\documentclass[utf8]{FrontiersinHarvard} 

\usepackage{url,hyperref,lineno,microtype,subcaption}
\usepackage[onehalfspacing]{setspace}
\usepackage{ulem}
\usepackage{makecell, multirow}
\usepackage{tabularx}
\usepackage{graphicx}
\usepackage{longtable}
\usepackage{supertabular}

\def\keyFont{\fontsize{8}{11}\helveticabold }
\def\firstAuthorLast{Lekshmi {et~al.}} 
\def\Authors{B Lekshmi\,$^{1,3,*}$, Kiran Jain\,$^{2}$ , Rudolf W. Komm\,$^{2}$ and Dibyendu Nandy\,$^{3,4}$}
\def\Address{$^{1}$Max Planck Institute for Solar System Research, Göttingen 37077, Germany\\
$^{2}$National Solar Observatory, Boulder, CO 80303, USA \\
$^{3}$Center of Excellence in Space Sciences India, Indian Institute of Science Education and Research Kolkata, Mohanpur 741246, West Bengal, India \\
$^{4}$Department of Physical Sciences, Indian Institute of Science Education and Research Kolkata, Mohanpur 741246, West Bengal, India}

\def\corrAuthor{B Lekshmi}

\def\corrEmail{lekshmib@mps.mpg.de}

\begin{document}
\onecolumn
\firstpage{1}

\title[Sub-surface flows \& Flares]{Sub-surface Plasma Flows and the Flare Productivity of Solar Active Regions} 

\author[\firstAuthorLast ]{\Authors} 
\address{} 
\correspondance{} 

\extraAuth{}

\maketitle
\begin{abstract}
The extreme space weather conditions resulting from high energetic events likes solar flares and Coronal Mass Ejections (CMEs) demand for reliable space weather forecasting. The magnetic flux tubes while rising through the convection zone gets twisted by the turbulent plasma flows, energizing the system and resulting in flares. We investigate the relationship between the subsurface plasma flows associated with flaring active regions and their surface magnetic flux and current helicity. The near-surface horizontal velocities derived from the ring-diagram analysis of active region patches using Global Oscillation Network Group (GONG) Doppler velocity measurements are used to compute the fluid dynamics descriptors like vertical divergence, vorticity and kinetic helicity used in this work. The flaring active regions are observed to have large value of vertical vorticity and kinetic helicity. Also, the horizontal flow divergence, vorticity, flux, kinetic and current helicities are observed to be significantly correlated and evolve in phase with each other. We observe that the integrated values of the above flow and magnetic parameters observed one day prior to the flare are significantly correlated with the integrated flare intensity of the active region. Hence, we show that strong vorticity/kinetic helicities lead to larger active region twisting, presumably generating high-intensity flares. 
\tiny
 \keyFont{ \section{Keywords:} Sun, Helioseismology, Active regions, Plasma flows, Helicity, Flares} 
\end{abstract}

\section{INTRODUCTION}

Strong magnetic flux tubes rise from the base of the convection zone to the solar surface due to magnetic buoyancy \citep{1990A&A...239..326C, 1993ApJ...405..390F} forming active regions and spread out to the solar atmosphere due to the low gas pressure environment. These flux tubes get twisted by the turbulent plasma flows in the convection zone \citep{1998ApJ...507..417L, 2004ApJ...611.1149H, 2006JGRA..11112S01N}. Following the emergence of an active region, surface shearing motions and rotation may also energise and twist the overlying magnetic loops \citep{Kazachenko_2009, Vemareddy_2012}. The magnetic energy gets stored as field-aligned electric current in the flux tube \citep{2011LRSP....8....6S}. Hence, the twisted flux tubes gain high non-potentiality while rising to the solar atmosphere and leads to the build-up of large magnetic energy and helicity in the magnetic structure. The highly unstable flux tubes dissipate stored energy in the form of high energetic events likes Flares and CMEs. Apart from the above described surface dynamics, magnetic reconnection process in the solar corona can also result in these events.

The fluid dynamic descriptors like divergence, vorticity and kinetic helicity derived from subsurface flow measurements improve the ability to distinguish between flaring and non-flaring active regions \citep{2011SoPh..268..389K, 2009JGRA..114.6105K}. Flaring regions are characterized by large values of subsurface vorticity \citep{Mason_2006}. \citet{2005ApJ...630.1184K} has shown that the unsigned kinetic helicity density of flaring active regions correlates well with their corresponding total flare X-ray intensity. Studying the evolutionary curves of subsurface kinetic helicity, \citet{2014SoPh..289..493G} showed that a flare stronger than M5.0 can occur within 8 hours before or after a rise in amplitude.
\citet{2010ApJ...710L.121R} defined a parameter, Normalized Helicity Gradient Variance to study the connection between subsurface flows and flare productivity, and observed that this parameter increases two or three days prior to flare. However contradicting these observations, \citet{Braun_2016} reported that no visible precursor associated with high-intensity flares were observed using the descriptors obtained using helioseismic holography technique.

The twist of magnetic field lines indicates stressed, non-potential flux system and is a manifestation of magnetic helicity \citep{PhysRevLett.33.1139, 2008JApA...29...49P}. \citet{2005ApJ...629.1135H} reported that flare occurs around regions with a high gradient in the magnetic twist and observations show that variance of twist distribution tends to decrease for more flare productive active regions \citep{2003ApJ...597L..73N, 2008ASPC..383..201N}. A more recent study \citep{2022arXiv220405910S} using machine learning models has also shown that total current helicity, unsigned flux and total absolute twist of active regions are good indicators of flare productivity. \citet{2010ApJ...710L.121R} proposed that the magnetic field lines form an unstable configuration due to the subsurface rotational kinetic energy, resulting in explosive reconnection and flare. Also, the photospheric kinetic helicity and current helicity of active regions are observed to be correlated \citep{Gao_2012}. These observations indicate that the twisting of field lines beneath the active region surface plays an important role in defining the current helicity distribution observed on the photosphere. 

The extreme space weather conditions resulting from solar flares and CMEs make it essential to have a proper understanding of the physical processes responsible for these events and articulate methods to predict their occurrence and strength. The necessity to develop tools to predict the occurrence of solar flares calls for further studies exploring the empirical connection between subsurface flows and the flare productivity of active regions. 
In this article, we study the plasma flow and magnetic field properties of flaring active regions. The plasma flow velocities obtained from the helioseismic measurements are used to calculate the subsurface fluid dynamic descriptors around selected active regions. The temporal evolution of subsurface flow divergence, vorticity, magnetic flux, kinetic and current helicities of these regions and their correlation with flare intensity are investigated. Our study provides observational evidence linking active region parameters to the generation of high-intensity flares. 

\section{ DATA and METHODS}
\begin{figure}[h!]
\begin{center}
\includegraphics[width = \textwidth]{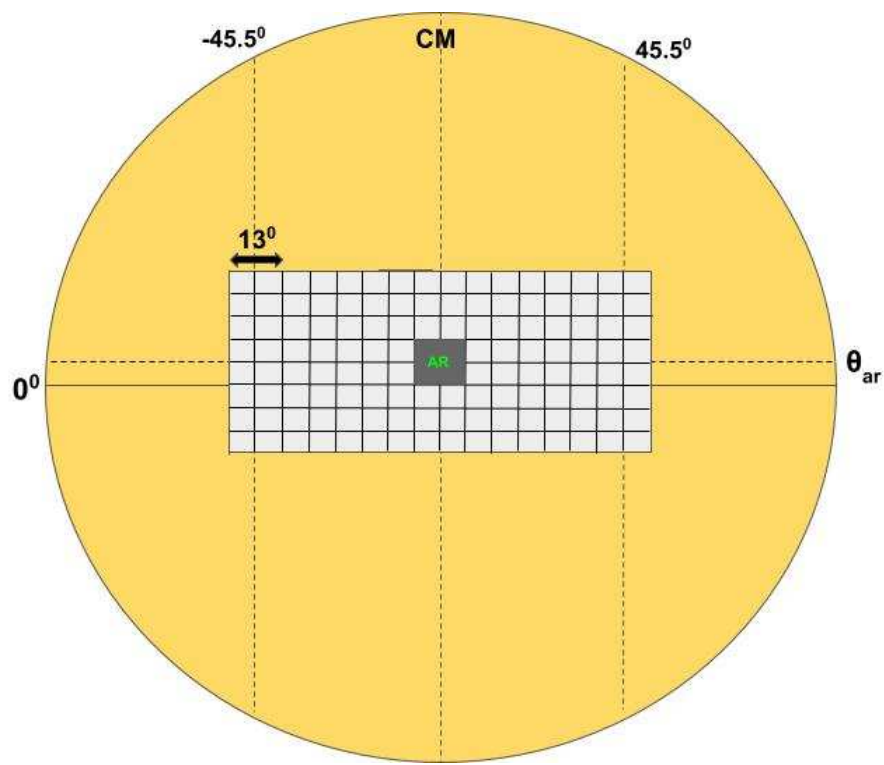}
\end{center}
\caption{A cartoon representing the selection of patches for ring diagram analysis. The active region latitude is represented as $\theta_{ar}$. Each patch is of size $13^{\circ}$ and there are 15 overlapping patches across longitude centered between $\pm 45.5^{\circ}$ from the central meridian (CM). Corresponding to each longitude, there are 7 overlapping patches across latitude with the middle patches centered at $\theta_{ar}$. The cartoon here represents the patch selection on the fourth day of tracking, when the active region (grey shaded area) patch is along the central meridian. (Grid sizes are not up to scale)}\label{fig:1}
\end{figure}

For our analysis, we select 18 isolated active regions which produce high-intensity (M \& X class) flares in solar cycle 24. The tracking period, observed flares and maximum flux (during the tracking period) of these active regions are listed in Table \ref{table:1}. A control group of 12 non-flaring active regions are also selected. All the active regions are tracked for 7 days with the central meridian (CM) distance of the patch center constrained between $\pm 45.5^{\circ}$ during the tracking period to minimize systematic effects. We select 105 overlapping Global Oscillation Network Group \citep[GONG;][]{Harvey1996, Jain2021} Dopplergram patches of size $13^{\circ} \times 13^{\circ}$ on each day. The patches are selected such that there are 15 overlapping patches centered between longitude $\pm 45.5^{\circ}$ and active region latitude ($\theta_{ar}$). Also, 3 overlapping patches are selected on either side of each longitudinal patch. A cartoon representing the patch selection is shown in Fig \ref{fig:1}. According to this selection criteria, the active region is centred at CM distance $-39^{\circ}$ and $39^{\circ}$ on the first and seventh day respectively. Hence we can define the longitudinal location of active region patch as $\varphi_{ar}$ = $-39^{\circ}$, $-26^{\circ}$, $-13^{\circ}$, $0^{\circ}$, $13^{\circ}$, $26^{\circ}$ and $39^{\circ}$ respectively on the 7 days of tracking. Ring-diagram analysis  is performed on these customized active region patches to obtain the horizontal flow ($v_{h}$) velocities \citep{1988ApJ...333..996H, 2005LRSP....2....6G}. In this study, selected $13^{\circ} \times 13^{\circ}$ patches are tracked for 1440 minutes with 1 minute cadence Dopplergrams using the surface rotation rate \citep{Snodgrass_1984}. Each tracked patch is apodized with a circular function and finally stacked in time to form a data cube. The 3-dimensional Fourier transform of this data cube in both spatial and temporal directions gives the 3-dimensional power spectra  with power concentrated at constant temporal frequencies. These frequencies get shifted in the presence of horizontal flows. The spectra are fitted by Lorentzian profile to obtain the frequency shifts, depth averaged horizontal flow components and other parameters. The flow components from the fit are inverted using regularized least square method to obtain the horizontal flow components at different depths \citep{Jain_2012, Jain_2015}. The measured horizontal flow velocities are interpolated for 33 selected depths between 0 - 15.5 Mm by linear interpolation. The temporal mean of the horizontal flow components (zonal and meridional) of each patch is calculated, and is subtracted from all the individual patches to remove systematic effects. Velocities corresponding to the patch containing the active region are excluded while calculating the mean. 

We compute plasma flow parameters such as horizontal flow divergence ($\nabla_r$), vorticity ($\omega_{r}$), and kinetic helicity ($K_{r}$) of each patch as given below \citep{2019ApJ...887..192K}:
\begin{equation}
   \nabla_r = \bigtriangledown . v_{h} = \frac{1}{r sin\theta}\Big( \frac{\partial v_{\varphi}}{\partial \varphi}\Big) + \frac{1}{r sin\theta}\Big( \frac{\partial}{\partial \theta}( sin\theta v_{\theta}) \Big),
\end{equation}
\begin{equation}
 \omega_{r} =  \bigtriangledown \times v_{h} = \frac{1}{r sin\theta}\Big( \frac{\partial}{\partial \theta} (sin\theta v_{\varphi})\Big) - \frac{1}{r sin\theta}\Big(\frac{\partial v_{\theta}}{\partial \phi}\big), 
\end{equation}
where $\theta$ and $\varphi$ are the colatitude and longitude corresponding to the patch center and, $v_{\varphi}$ and $v_{\theta}$ represents the zonal and meridional flows.The kinetic helicity at different depths is given by
\begin{equation}
    \mathrm{K}_{r} = \int^{d}_{R_{\odot}}k_{r}(r) A(r) dr,
    \label{eqn:kr}
\end{equation}
where $k_{r} = \omega_{r} . v_{r}$, d is the depth, $v_{r}$ is the radial velocity and A is the area of the patch. The radial velocity,
\begin{equation}
    v_{r}(d) = -\frac{1}{\rho(d)}\int^{d}_{R_{\odot}}\rho \bigtriangledown.v_{h} dr + v_{r}(R_{\odot})\frac{\rho(R_{\odot})}{\rho(d)}.
\end{equation}
The density $\rho(d)$ at each depth $d$ is calculated from a solar model - Model S \citep{1998SSRv...85...19C} and $v_{r}$ on the solar surface is considered to be zero.

The surface magnetic flux ($\Phi$) and total unsigned current helicity ($\mathrm{J_z^{tot}}$), where $\mathrm{J_z = (\nabla \times B_z) . B_z}$, (magnetic field - $\mathrm{B_z}$) of the selected flaring and non-flaring active regions observed at the tracking time are obtained from HMI-SHARP data \citep{2014SoPh..289.3549B}. The $\Phi$ and $\mathrm{J_z^{tot}}$ are in unis of Maxwell and $\mathrm{G^2/m}$ respectively. The intensities of all the flares produced by the selected active regions during the tracking period is obtained from the Geostationary Operational Environmental Satellite (\href{https://www.ngdc.noaa.gov/stp/space-weather/solar-data/solar-features/solar-flares/x-rays/goes/xrs/}{GOES}) data. The intensities of all the flares corresponding to each active region on each day of observation are integrated to obtain a quantity named as Integrated Flare Intensity (FI) in units of $Wm^{-2}$ (see Table \ref{table:2}).

\section{RESULTS}
\begin{figure}[h!]
\begin{center}
\includegraphics[width = \textwidth]{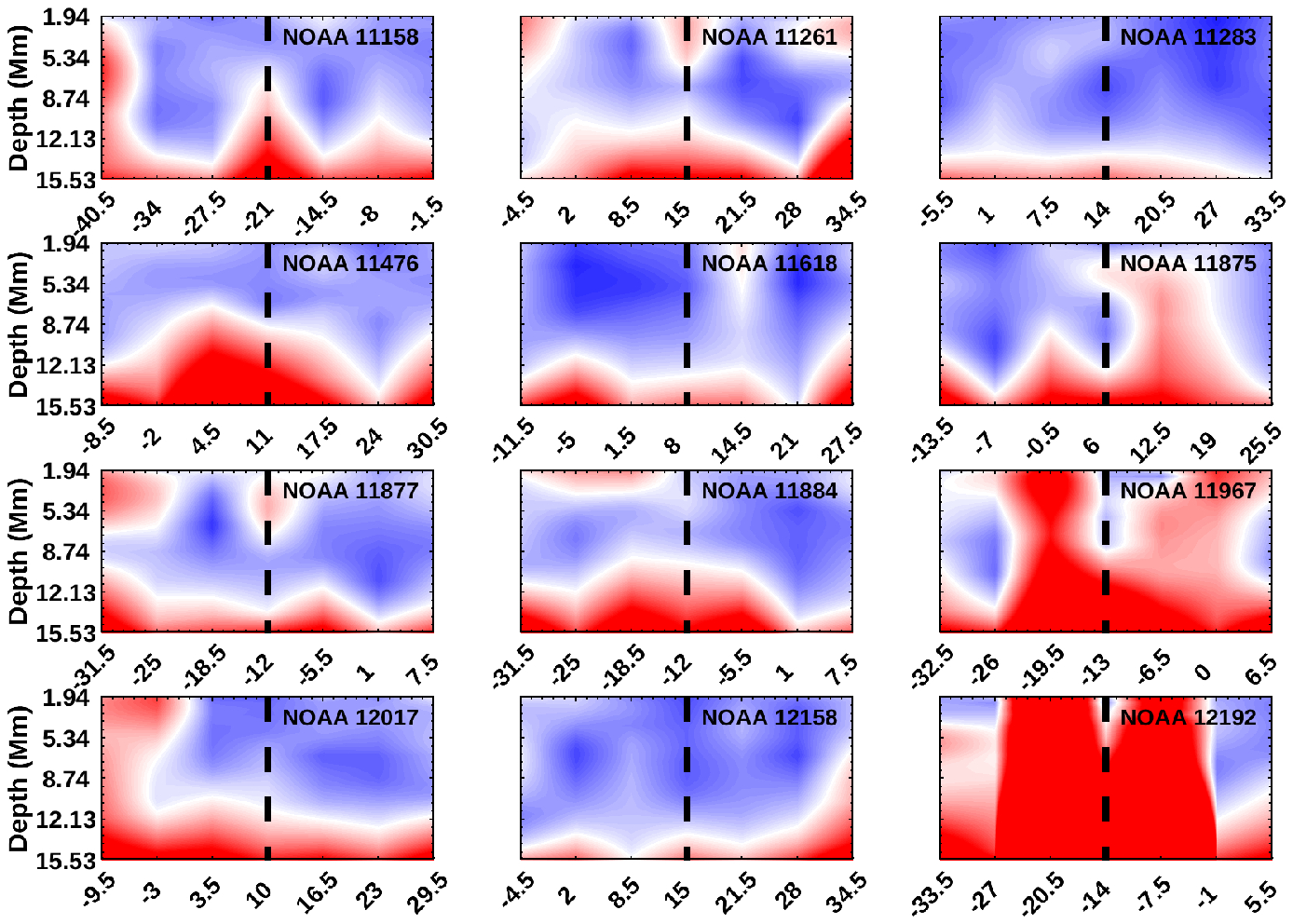}\\
\vspace{-0.15cm}
\includegraphics[width = \textwidth]{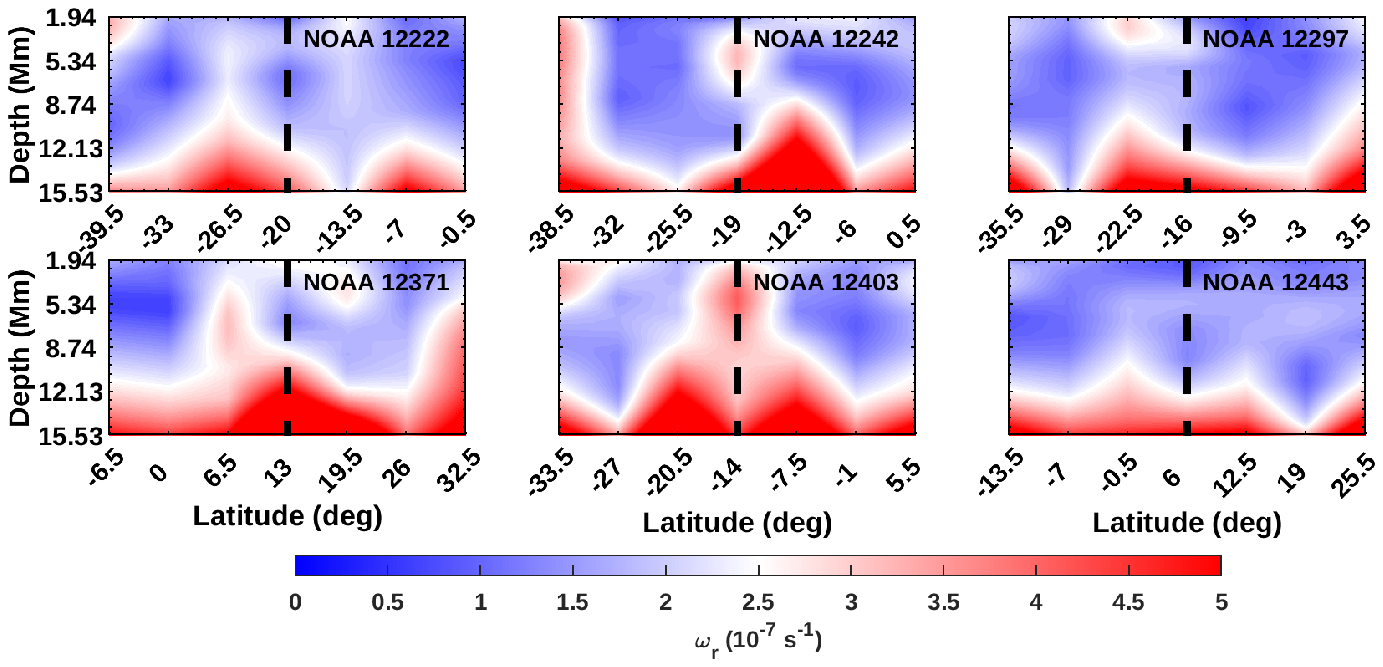}
\end{center}
\caption{Temporal mean of absolute vertical vorticity, $\omega_r$ corresponding to the 18 flaring active regions and neighboring patches at the active region longitude, $\varphi_{ar}$ as a function of latitude and depth. The latitude of active region is represented by dashed solid line.}\label{fig:2}
\end{figure}

\begin{figure}[h!]
\begin{center}
\includegraphics[width = 1\textwidth]{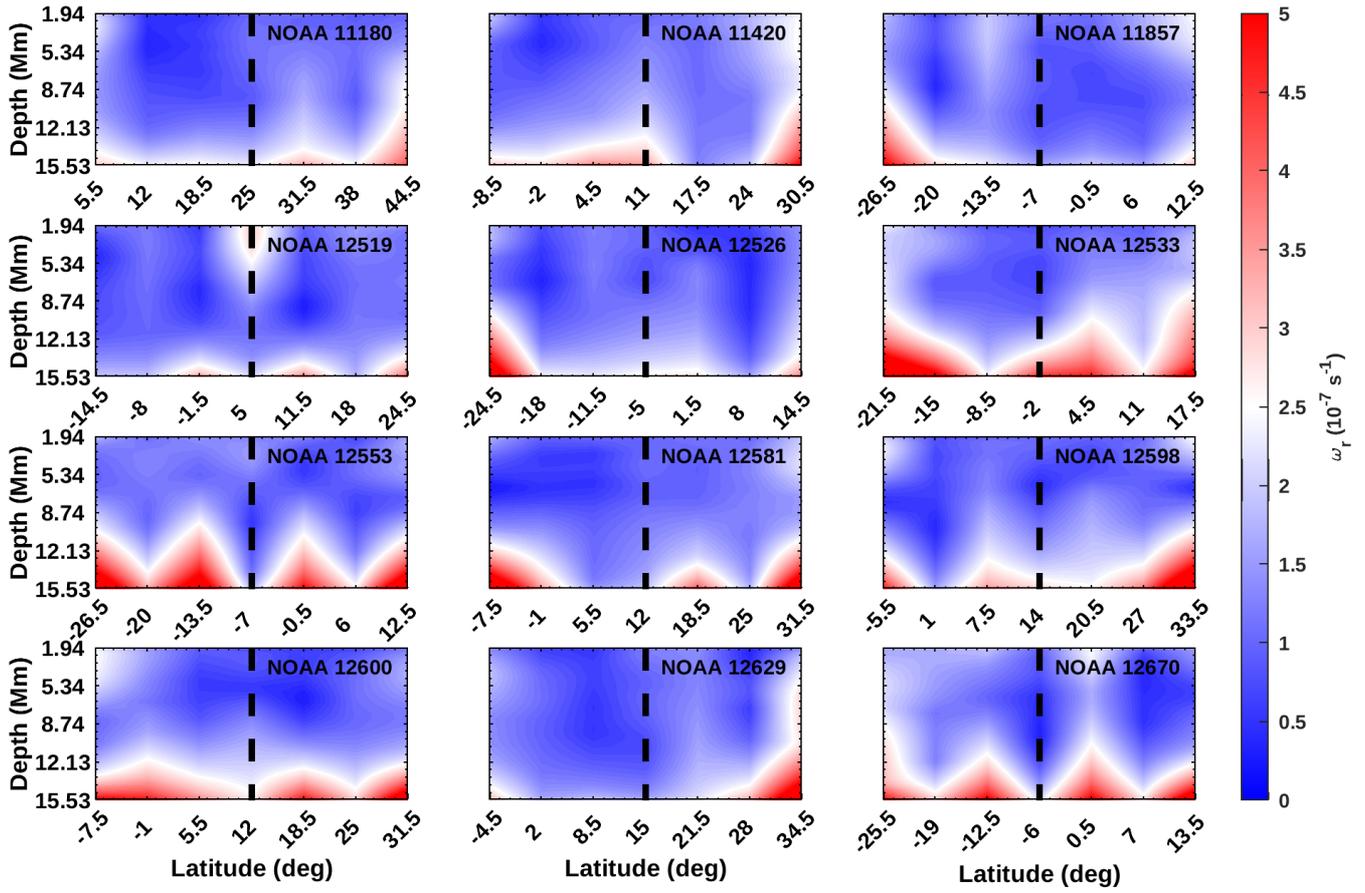}
\end{center}
\caption{Same as Fig \ref{fig:2} but for 12 non-flaring active regions.}\label{fig:3}
\end{figure}

We compute the absolute values of horizontal flow divergence, vertical vorticity and kinetic helicity of all the tracked patches using systematic corrected flow velocities. The temporal mean (over the 7 days of tracking) of vertical vorticity, $\omega_r$, at each depth, corresponding to the flaring active region and its neighboring latitude patches at $\varphi_{ar}$ are plotted as a function of latitude and depth in Fig \ref{fig:2}. The latitude of active region, $\theta_{ar}$ is represented by black dashed line.
The mean vorticity of active region patches and neighboring latitudes are observed to have large values and are strongest at deeper layers. It can be observed from Fig \ref{fig:2} that regions like 11967, 12192 and 12403 have large values of vorticities over all the depths considered. These regions have produced numerous X and M class flares during the tracking period and have large values of magnetic flux ($\mathrm{> 6 \times 10^{22} Mx}$; see Table \ref{table:1}). However, regions like 11283, 11618 and 12158 have strong vorticities concentrated at deeper layers and have weaker magnetic flux. These regions produce fewer X and M class flares compared to the above mentioned case. This indicates that regions which produce more number of high intensity flares have large vorticities ($\mathrm{> 2 \times 10^{-7} s^{-1}}$) spanning from deeper layers till the surface. The mean vorticity of non-flaring active region patches along with that of neighboring patches are plotted in Fig \ref{fig:3}. The vorticity at the active region locations (dashed lines) are smaller in amplitude as compared to that of flaring regions. For comparison, the color bars of both the figures are saturated to vorticity $5 \times 10^{-7} \mathrm{s^{-1}}$.

\begin{figure}[h!]
\begin{center}
\includegraphics[width = \textwidth]{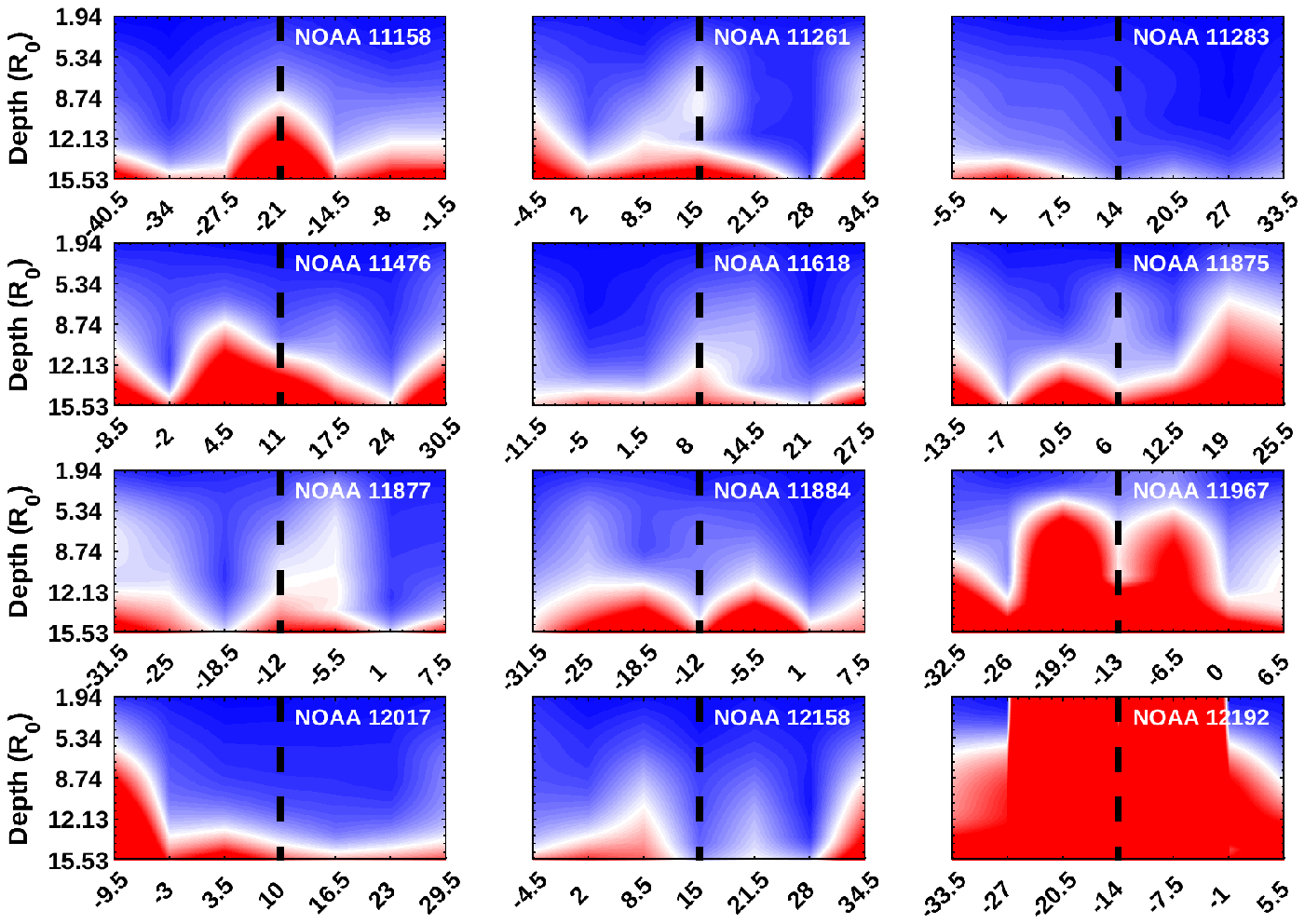}\\
\vspace{-0.15cm}
\includegraphics[width = \textwidth]{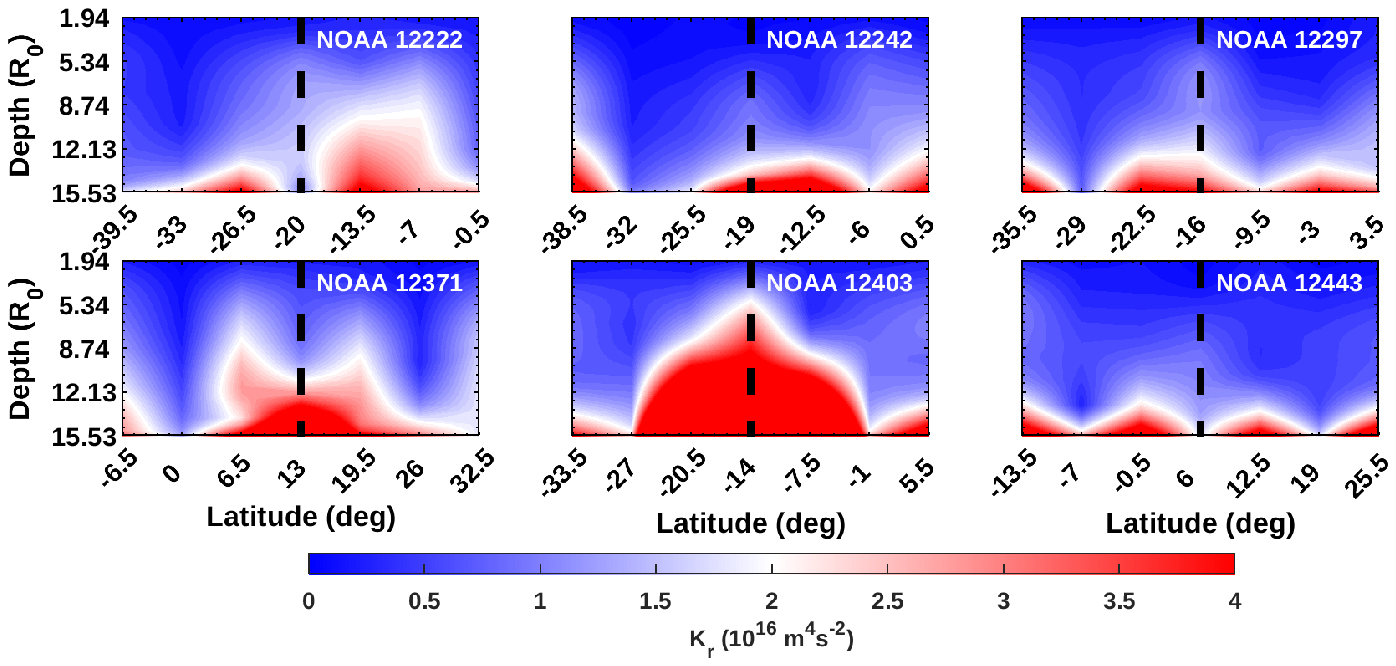}
\end{center}
\caption{Temporal mean of absolute kinetic helicity, $\mathrm{K}_r$ corresponding to the flaring active regions and neighboring patches at $\varphi_{ar}$ as a function of latitude and depth. Dashed lines represent the active region latitude.} \label{fig:4}
\end{figure}

\begin{figure}[h!]
\begin{center}
\includegraphics[width = \textwidth]{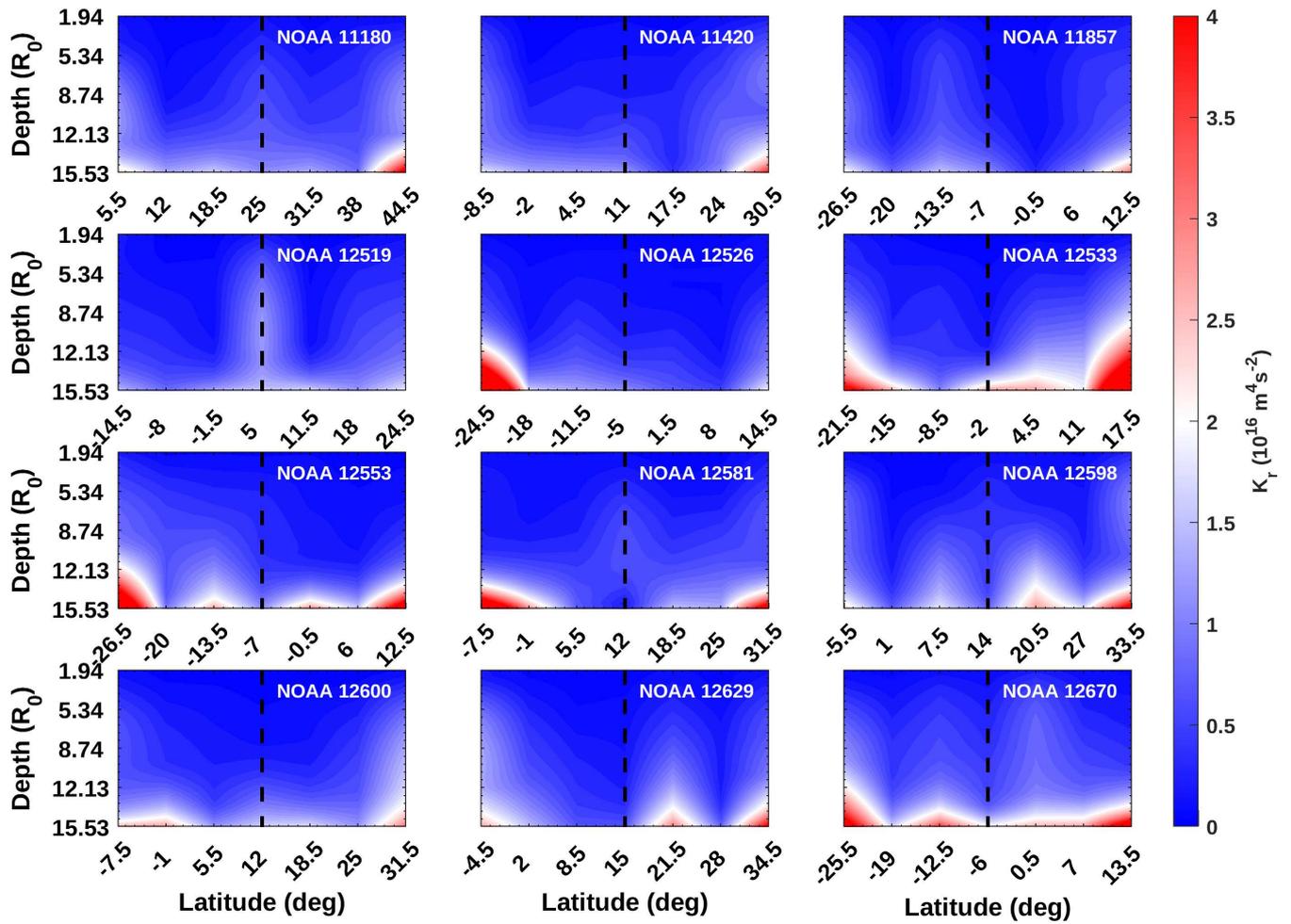}
\end{center}
\caption{Same as Fig \ref{fig:4} but for non-flaring active regions and neighboring patches.}\label{fig:5}
\end{figure}

The temporal mean of kinetic helicity, $\mathrm{K}_r$, of the flaring and non-flaring active regions along with the helicities of neighboring latitudes at $\varphi_{ar}$ are plotted in Fig \ref{fig:4} and \ref{fig:5}, respectively. The kinetic helicity of flaring active regions and neighboring latitudes are larger in amplitude ($\mathrm{> 10^{15} m^4s^{-2}}$ over all depths) with respect to non-flaring regions, similar to that of vorticity. Also, the $\mathrm{K}_r$ of flaring active regions have larger values ($\mathrm{> 10^{16} m^4s^{-2}}$) at deeper layers. The colorbars are saturated to $\mathrm{4 \times 10^{16} m^4s^{-2}}$.

\begin{figure}[h!]
\begin{center}
\includegraphics[width = \textwidth]{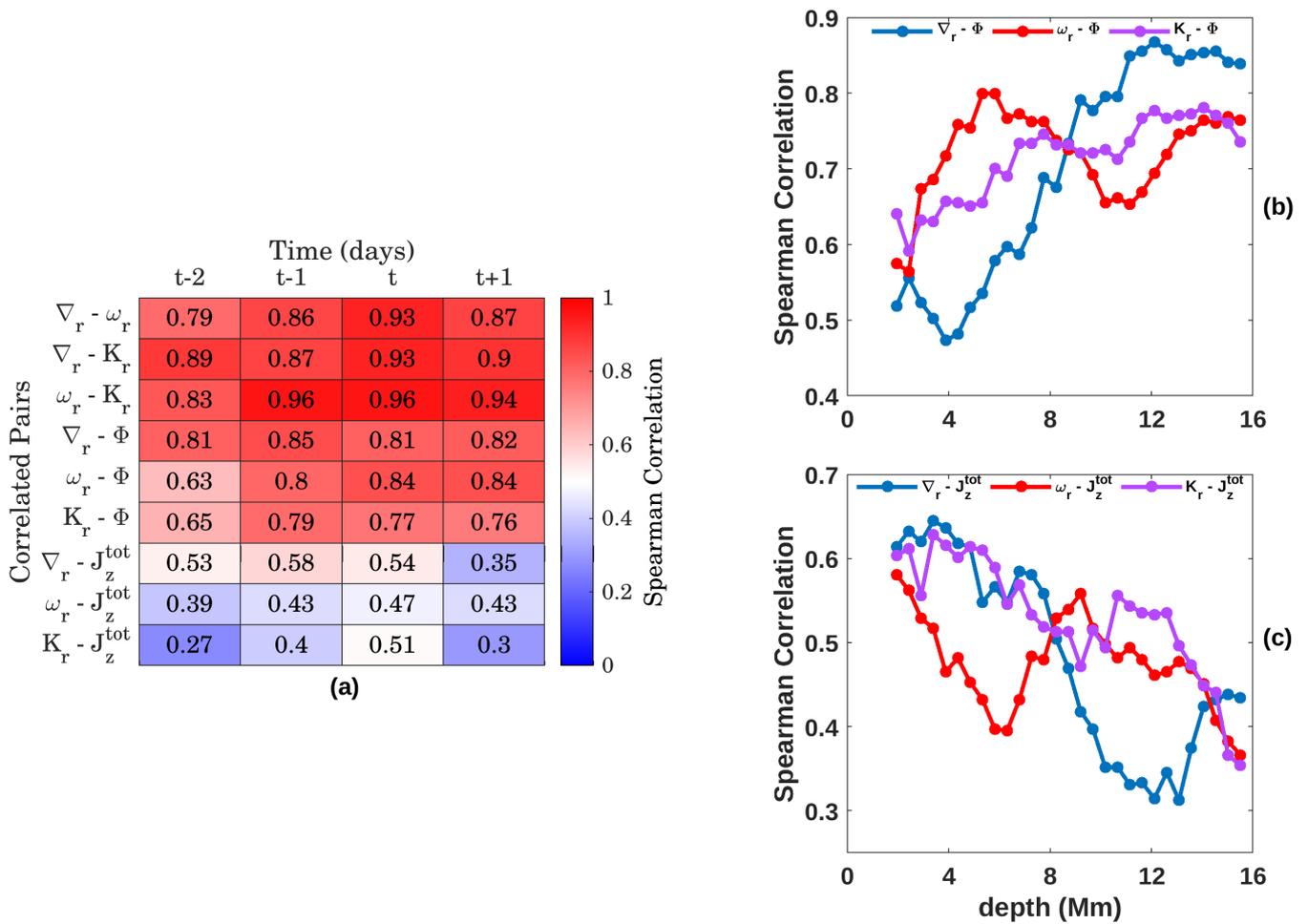}
\end{center}
\caption{(a) Spearman correlation coefficients between the integrated  horizontal flow divergence ($\nabla_r$), vertical vorticity ($\omega_r$), kinetic helicity ($\mathrm{K}_r$), surface magnetic field ($\Phi$) and photospheric current helicity ($\mathrm{J_z^{tot}}$) of active regions two days prior to (t-2), one day prior to (t-1), on the day (t) and one day after (t+1) the flare. Color bar represents the strength of correlations and all the values $>0.45$ have significance $>95\%$. The flow parameters are averaged over 2 - 15.5 Mm. (b) Spearman correlation between the integrated flow parameters on the day of flare with the corresponding surface magnetic flux. (c) Spearman correlation between the integrated flow parameters on the day of flare and  photospheric current helicity. All the correlation coefficients in (b) and (c) have significance $> 95\%$.}\label{fig:6}
\end{figure}

The active regions selected for our analysis have flared multiple times during the tracking period. The days (t) on which flares are observed and the corresponding integrated flare intensities are given in Table \ref{table:2}. In order to understand the inter-relationship between plasma flow and magnetic properties of flaring active regions, we integrate the $\nabla_r$, $\omega_r$, $\mathrm{K}_r$, $\Phi$ and $\mathrm{J_z^{tot}}$  of the active region patches on two days prior (t-2), one day prior (t-1), on the day (t) and one day post (t+1) all the flares observed during the tracking period. The $\nabla_r$, $\omega_r$, $\mathrm{K}_r$ are averaged over depths 2 - 15.5 Mm. A correlation analysis is performed between the integrated values of parameters and the Spearman correlation values are shown in Fig \ref{fig:6}(a). All the correlations have a significance $> 95\%$ except the one's with correlations $<0.45$. All the flow parameters are strongly correlated with each other and with magnetic flux on all the 4 time periods considered here. However, the flow parameters show highest correlation with $\mathrm{J_z^{tot}}$ on the day of flare. The strong correlations obtained between the parameters at different time stamps with respect to the flaring time, indicate that they evolve in phase with each other. Also, it is interesting to note that the inter relationship between the flow and magnetic properties are strongest on the day of flare and one day prior to it. As a control experiment, the sum of parameters obtained for non-flaring active regions over all days of observation are computed and a correlation analysis is performed between these parameters. We cannot find any significant correlation between the parameters as the median value of the correlations obtained is 0.2  with significance $< 70 \%$.

The relationship of the flow parameters at different depths with the current helicity and magnetic flux measurements at the surface are also investigated. Fig \ref{fig:6} (b) and (c) represents the Spearman correlation coefficients computed between the integrated flow parameters at different depths on the day of flare with $\Phi$ and $\mathrm{J_z^{tot}}$ respectively. The highest correlations between the parameters were observed on the day of flare, hence we consider them for the depth analysis. It is observed that the correlations of the parameters with magnetic flux increases with depth whereas, it is strongest at the shallower depths in the case of current helicity. All the correlations reported here have a significance $> 95\%$. We also check whether the the subsurface kinetic helicity precedes the current helicity observed on the day of flare. No significant correlation is obtained between these two parameters (significance $<70\%$ over all depths) implying the current helicity evolves in phase with the kinetic helicity even at deeper layers, similar to that reported by \cite{Gao_2012} for shallow depths (0-1 Mm). We also notice that the vorticity close to the surface (2Mm) shows significantly higher correlation ($> 97\%$; $> 0.5$) with that at deeper layers  ($\approx$ 7 - 8Mm), observed on the same and day prior. More detailed analysis on the evolution of surface and subsurface vorticities and their inter relationship might help to explain the role of photospheric inflows in magnetic field twisting and buildup  of current helicity. Also, the role of subsurface kinetic helicity in the build up current helicity and their in-phase evolution. But this needs a comprehensive analysis of active region flows at the surface and subsurface layers which provides scope for another study using the improved spatial-resolution Dopplergrams.

\begin{figure}[h!]
\begin{center}
\includegraphics[width = \textwidth]{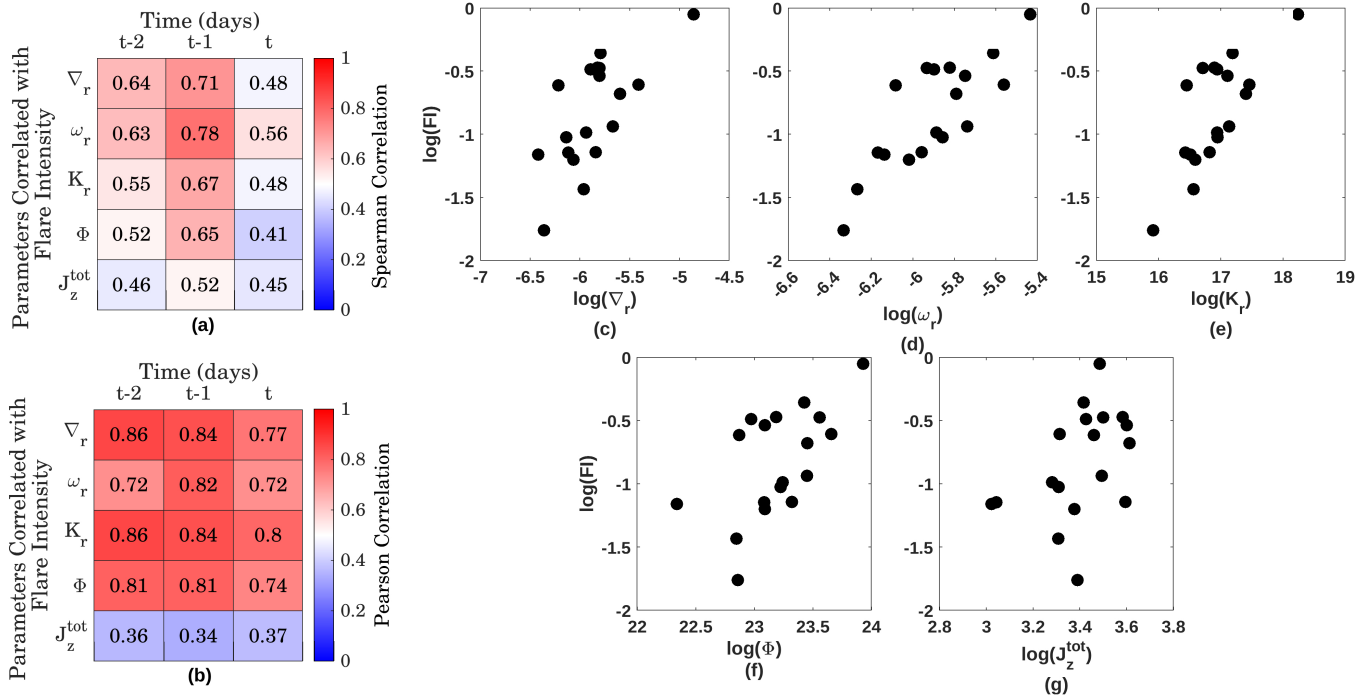}
\end{center}
\caption{The integrated values of active region horizontal flow divergence ($\nabla_r$), vertical vorticity ($\omega_r$), kinetic helicity ($\mathrm{K}_r$), surface magnetic field ($\Phi$)  and photospheric current helicity ($\mathrm{J_z^{tot}}$) obtained on t-2, t-1, and t days of flare are correlated with integrated flare intensity (FI). The flow parameters are averaged over 2 - 15.5 Mm. (a) Mean of Spearman correlation coefficients computed by removing the parameters corresponding to all unique permutations of 0 - 3 elements from the 18 flaring active regions. (b) Mean Pearson correlation coefficients computed as in (a). Color bar represents the strength of correlations and the values $> 0.4$ have significance $> 95\%$.  Scatter plot of integrated flare intensity  as a function of integrated (c) horizontal flow divergence, (d) vertical vorticity, (e) kinetic helicity, (f) surface magnetic field and (g) photospheric current helicity observed one day prior to flare in logarithmic scale.}\label{fig:7}
\end{figure}

Subsequently, we study the role of fluid and magnetic parameters of active regions with the intensity of flares they produce. For this, we correlate the integrated $\nabla_r$, $\omega_r$, $\mathrm{K}_r$, $\Phi$ and $\mathrm{J_z^{tot}}$  on t-2, t-1 and t days with the integrated FI of each active region. The flow parameters are averaged over depths 2 - 15.5 Mm. It is observed that all the parameters computed shows strong correlations with flare intensity. However, to confirm the reliability on the correlations and see whether they are biased by any particular set of active region samples, we consider all unique permutations of 0 - 3 elements from the set of 18 active regions we are analysing. The parameters corresponding to each of the permuted sets are removed and the correlations of the remaining samples are computed. Figure \ref{fig:7} (a) and (b) represents the mean of Spearman and Pearson correlation coefficients obtained by removing active regions as explained above. The color gradient represents the strength of correlations. It can be observed that significant correlations are observed on the three days with the strongest correlations on the day before the flare. All the correlation values $> 0.4$ have significance $> 95\%$. Fig \ref{fig:7} (c - g) represents the scatter plots of total FI as a function of the total $\nabla_r$, $\omega_r$, $\mathrm{K}_r$, $\mathrm{J_z^{tot}}$ and $\Phi$ observed one day prior to the flare, plotted in logarithmic scale. The plot and Pearson correlation coefficients given in the table indicate that the parameters except $\mathrm{J_z^{tot}}$ have a linear relationship with the flare intensities. 

\section{ DISCUSSION}

\citet{2009JGRA..114.6105K} reported a linear relationship between surface magnetic flux and the subsurface vorticity of active regions and observed that regions which produce high intensity flares are characterized by large values of flux ($\mathrm{> 80 Gauss}$) and vorticity ($\mathrm{> 0.86 \times 10^{-5} s^{-1}}$). \citet{Braun_2016}  also reported results which qualitatively agrees with the above observation. We analyse the vertical vorticity and kinetic helicity of 18 flaring and 12 non-flaring active regions and observe that flaring active regions are characterized by large values of vorticity and helicity compared to its non-flaring counterpart. Also, vorticity and kinetic helicity strength of the flaring regions increase with depth. Our results also agree qualitatively with the previous studies.

The subsurface flow parameters of flaring active regions are significantly correlated with each other and significant correlations obtained on the days prior to, on the day and post flare indicate that the parameters evolve in phase with each other. The same is observed between the flow parameters and surface magnetic flux. Plasma flows converge towards active regions \citep{2001IAUS..203..189G, 2004ApJ...603..776Z} due to the temperature perturbations resulting from the presence of strong magnetic field \citep{1981A&A....94L..17S,2003SoPh..213....1S}. The flow convergence increases with magnetic flux, increasing the strength of divergence and vorticity around the active region and hence the kinetic helicity(Eq \ref{eqn:kr}). The flow parameters show significant correlation with photospheric current helicity one day prior to and on the day of flare. The rise in divergence and vorticity around active regions strengthens the kinetic helicity which results in the twisting of active region flux tubes and this is known as $\Sigma$-effect \citep{1998ApJ...507..417L}. The kinetic and current helicities evolve in phase indicating that the twisting of flux tubes below the active region results in the build up of photospheric current helicity \citep{Gao_2012}. We observe that the correlation between the  
flow parameters and flux increases with depth but incase of current helicity maximum correlations are observed at shallower depths. 

In the presence of strong magnetic field, the subsurface divergence, vorticity and kinetic helicity of the active regions increases in strength. The resultant shearing motion twists the magnetic flux tubes beneath the surface, which are frozen in the plasma and energises them. The energized sub-surface flux systems are prone to be more flare productive \citep{2003ApJ...597L..73N}. Studies have shown that kinetic helicity/vorticity measurements of active regions show signatures of an upcoming flare, few days to hours prior to the event \citep{2010ApJ...710L.121R, 2014SoPh..289..493G}. We observe that the flow parameters, flux and current helicity one day prior to the flare shows strong correlation with the intensity of the flares generated. This indicates that these parameters one day prior to the flare might be able to provide information about the strength of the flares that an active region can produce during its passage through the solar disc. However, more detailed studies on temporal evolution of active region parameters are required to explore the possibilities of using them to predict the strength of individual flares.

\section*{Conflict of Interest Statement}

The authors declare that the research was conducted in the absence of any commercial or financial relationships that could be construed as a potential conflict of interest.

\section*{Author Contributions}

All the authors contributed to the conceivement of the presented idea. BL, KJ and RW designed the analysis. Ring diagram velocity measurements of customised patches were done by KJ. BL performed the analysis and wrote the manuscript. RW, KJ and DN along with BL contributed to the interpretation of results and in the preparation of final manuscript. 

\section*{Funding}
This work was supported by the Council of Scientific and Industrial Research and Ministry of Human Resource Development, Government of India. BL acknowledges Max Planck Society for the funding. KJ and RWK acknowledge the support from NASA grants 80NSSC18K1206, 80NSSC19K0261, 80NSSC20K0194, and 80NSSC21K0735 to the National Solar Observatory and by NASA grant NNH18ZDA001N-DRIVE to Stanford University.

\section*{Acknowledgments}
This work utilises GONG data obtained by the NSO Integrated Synoptic Program, managed by the National Solar Observatory, which is operated by the Association of Universities for Research in Astronomy (AURA), Inc. under a cooperative agreement with the National Science Foundation and with contribution from the National Oceanic and Atmospheric Administration. The GONG network of instruments is hosted by the Big Bear Solar Observatory, High Altitude Observatory, Learmonth Solar Observatory, Udaipur Solar Observatory, Instituto de Astrof\'{\i}sica de Canarias, and Cerro Tololo Interamerican Observatory. The data used here are courtesy of NASA/SDO and the HMI Science Team. We acknowledge the use of Geostationary Operational Environmental Satellite (\href{https://www.ngdc.noaa.gov/stp/space-weather/solar-data/solar-features/solar-flares/x-rays/goes/xrs/}{GOES}) data. The Center or Excellence in Space Sciences India at IISER Kolkata is funded by the Ministry of Education, Government of India.


\bibliographystyle{Frontiers-Harvard} 
\bibliography{references}

\newpage

\section*{Tables}

\begin{longtable}{|p{1cm}|p{3.7cm}|p{0.6cm}|p{0.6cm}|p{0.6cm}|p{0.6cm}|p{0.6cm}|p{0.6cm}|p{0.6cm}|p{2cm}|}
    \hline
\multirow{1}{*}{\makecell{NOAA}}
& \multirow{1}{*}{Tracking Time}
    & \multicolumn{7}{c|}{Flares produced}
     & \multirow{1}{*}{\makecell{$\Phi_{max}$\\ ($10^{22} \times \mathrm{Mx}$)}}    \\
    \cline{3-9}
      &  & $\mathrm{t_1}$& $\mathrm{t_2}$& $\mathrm{t_3}$& $\mathrm{t_4}$& $\mathrm{t_5}$& $\mathrm{t_6}$& $\mathrm{t_7}$ &      \\
    \hline
11158  & \makecell{2011-02-10 14:00:00 \\- 2011-02-17 13:59:00}  &             -  &             -  &         \makecell{1B\\2C}  &         \makecell{7C\\1M}  &      \makecell{3C\\1M \\1X}  &            8C  &        \makecell{14C\\1M}  &      3.064  \\
    \hline

11261  & \makecell{2011-07-29 00:00:00 \\- 2011-08-04 23:59:00}  &        \makecell{10C\\2M}  &         \makecell{3C\\1X}  &            2C  &         \makecell{5C\\1M}  &            6C  &            6C  &        \makecell{10C\\2M}  &      2.527  \\
    \hline

11283  & \makecell{2011-09-02 08:24:00 \\- 2011-09-09 08:23:00}  &            2C  &             -  &             -  &            1C  &            1C  &         \makecell{1C\\1M}  &         \makecell{5C\\1X}  &      2.618  \\
    \hline

11476  & \makecell{2012-05-07 23:24:00 \\- 2012-05-14 23:23:00}  &            2C  &         \makecell{5C\\1X}  &        \makecell{17C\\1M}  &        \makecell{14C\\2M}  &            7C  &         \makecell{2B\\1C}  &            3C  &      5.398  \\
    \hline

11618  & \makecell{2012-11-18 12:59:00 \\- 2012-11-25 12:58:00}  &             -  &         \makecell{1B\\2C}  &         \makecell{5C\\1M}  &            1C  &             -  &             -  &            1C  &      2.534  \\
    \hline


11875  & \makecell{2013-10-19 21:12:00 \\- 2013-10-26 21:11:00}  &             -  &         \makecell{5C\\1M}  &     \makecell{12C\\1M\\1X}  &         \makecell{3C\\1M}  &            5C  &            1C  &            4C  &      4.288  \\
    \hline

11877  & \makecell{2013-10-21 17:24:00 \\- 2013-10-28 17:23:00}  &             -  &            1B  &         \makecell{3C\\1M}  &             -  &            3C  &            2C  &            1C  &     3.910  \\
    \hline


11884  & \makecell{2013-10-29 12:23:00 \\- 2013-11-05 12:22:00}  &             -  &             -  &            3C  &            4C  &      \makecell{3C\\1M\\1X}  &            2C  &             -  &      4.458  \\
    \hline

11967  & \makecell{2014-01-30 20:23:00 \\- 2014-02-06 20:22:00}  &            1M  &            3C  &         \makecell{3C\\1X}  &            5C  &         \makecell{4C\\1M}  &            5C  &            2C  &      8.794  \\
    \hline

12017  & \makecell{2014-03-24 03:35:00 \\- 2014-03-31 03:34:00}  &             -  &             -  &             -  &             -  &      \makecell{1B\\2C\\1M}  &         \makecell{8C\\1M}  &             -  &      1.420  \\
    \hline

12158  & \makecell{2014-09-10 00:01:00 \\- 2014-09-17 00:00:00}  &            1C  &             -  &         \makecell{2C\\1M}  &            1C  &             -  &            3C  &             -  &      3.448  \\
    \hline

12192  & \makecell{2014-10-20 02:00:00 \\- 2014-10-27 01:59:00}  &     \makecell{10C\\1M\\1X}  &            5C  &            4C  &            4C  &            4C  &            1M  &         \makecell{7C\\1M}  &      18.409  \\
    \hline

12222  & \makecell{2014-11-28 18:35:00 \\- 2014-12-05 18:34:00}  &         \makecell{3C\\1M}  &            2C  &            1X  &            2C  &            4C  &            2C  &         \makecell{2C\\1M}  &      3.952 \\
    \hline

12242  & \makecell{2014-12-13 11:47:00 \\- 2014-12-20 11:46:00}  &             -  &      \makecell{1B\\4C\\1M}  &         \makecell{6C\\1M}  &             -  &         \makecell{3C\\1X}  &            4C  &         \makecell{7C\\1M}  &      10.176 \\
    \hline

12297  & \makecell{2015-03-09 16:35:00 \\- 2015-03-16 16:34:00}  &     \makecell{11C\\1M\\1X}  &        \makecell{16C\\1M}  &         \makecell{1C\\1M}  &            3C  &            8C  &         \makecell{2B\\8C}  &         \makecell{2B\\4C}  &      3.135  \\
    \hline

12371  & \makecell{2015-06-18 15:24:00 \\- 2015-06-25 15:23:00}  &            3C  &         \makecell{4C\\1M}  &            1X  &            2C  &            5C  &         \makecell{2C\\1M}  &            2C  &      5.339  \\
    \hline

12403  & \makecell{2015-08-20 07:01:00 \\- 2015-08-27 07:00:00} &         \makecell{8C\\1M}  &         \makecell{8C\\1X}  &         \makecell{9C\\1M}  &            4C  &        \makecell{17C\\2M}  &        \makecell{2B\\10C}  &         \makecell{2B\\7C}  &     6.716 \\
    \hline

12443  & \makecell{2015-10-31 21:00:00 \\- 2015-11-07 20:59:00}  &     \makecell{10C\\1M\\1X}  &            1C  &            1C  &            4C  &             -  &             -  &             -  &      4.617  \\
    \hline

    \cline{2-5}
    \caption{NOAA number, tracking time, number of flares  produced of each flare class (B,C,M and X) and the maximum unsigned flux of the active regions during the tracking period. The columns $\mathrm{t_1 - t_7}$ represents the 7 days of tracking. The number of flares corresponding to each flare class are indicated as prefix.}
\label{table:1}
    \end{longtable}

\begin{table}[]
    \centering
    \renewcommand{\arraystretch}{1.5}
    \begin{tabular}{|c|c|c|c|c|c|c|c|}
    \hline
    \multirow{1}{*}{\makecell{NOAA}}
    & \multicolumn{7}{c|}{Integrated Flare Intensity ($\mathrm{Wm^{-2}}$)} \\
    
     \cline{2-8}
        & $\mathrm{t_1}$& $\mathrm{t_2}$& $\mathrm{t_3}$& $\mathrm{t_4}$& $\mathrm{t_5}$& $\mathrm{t_6}$& $\mathrm{t_7}$ \\
\hline
11158 & - & - & 0.007 & 0.080 & 0.174 & 0.031 & 0.033\\
\hline
11261 & 0.010 & 0.021 & 0.004 & 0.008 & 0.042 & 0.149 & 0.067\\
\hline
11283 & 0.001 & - & - & 0.054 & 0.058 & 0.070 & 0.062\\
\hline
11476 & 0.005 & 0.028 & 0.041 & 0.026 & 0.006 & 0.009 & 0.005\\
\hline
11618 & - & 0.011 & 0.025 & 0.0001 & - & - & 0.001\\
\hline
11875 & - & 0.011 & 0.037 & 0.017 & 0.027 & 0.003 & 0.007\\
\hline
11877 & - & 0.001 & 0.059 & - & 0.009 & 0.004 & 0.021\\
\hline
11884 & - & - & 0.005 & 0.025 & 0.027 & 0.007 & -\\
\hline
11967 & 0.005 & 0.038 & 0.069 & 0.014 & 0.096 & 0.021 & 0.009\\
\hline
12017 & - & - & - & - & 0.016 & 0.053 & -\\
\hline
12158 & 0.38 & - & 0.010 & 0.0004 & - & 0.006 & -\\
\hline
12192 & 0.183 & 0.013 & 0.357 & 0.022 & 0.025 & 0.002 & 0.470\\
\hline
12222 & 0.0197 & 0.002 & 0.023 & 0.011 & 0.015 & 0.019 & 0.003\\
\hline
12242 & - & 0.016 & 0.008 & - & 0.018 & 0.013 & 0.279\\
\hline
12297 & 0.029 & 0.186 & 0.027 & 0.006 & 0.023 & 0.067 & 0.027\\
\hline
12371 & 0.014 & 0.025 & 0.032 & 0.009 & 0.196 & 0.008 & 0.17068\\
\hline
12403 & 0.031 & 0.064 & 0.039 & 0.004 & 0.025 & 0.007 & 0.071\\
\hline
12443 & 0.040 & 0.008 & 0.003 & 0.060 & - & - & -\\
\hline
    \end{tabular}
    \caption{Integrated intensity of all the flares produced by the active regions (see Table \ref{table:1}) on each day of tracking.}
    \label{table:2}
\end{table}

\section*{Figure captions}


\end{document}